\documentclass[runningheads]{llncs}
\usepackage[T1]{fontenc}
\usepackage{graphicx}

\usepackage{booktabs}       
\usepackage{amsfonts}      
\usepackage{nicefrac}       
\usepackage{microtype}      
\usepackage{algorithm}
\usepackage{algpseudocode}
\usepackage{appendix}
\usepackage{multirow}
\usepackage{caption}
\usepackage{babel}
\usepackage{makecell}
\usepackage{enumitem}
\usepackage[misc]{ifsym}

\begin{document}

\title{ProteinEngine: Empower LLM with Domain Knowledge for Protein Engineering}
\author{Yiqing Shen\inst{1}
\and
Outongyi Lv\inst{2}\and
Houying Zhu\inst{3}\and
Yu Guang Wang\inst{2}\textsuperscript{(\Letter)}
}
\authorrunning{Y. Shen et al.}
%
\institute{Department of Computer Science, Johns Hopkins University, USA\\
\email{yshen92@jhu.edu}, \email{yuguang.wang@sjtu.edu.cn}\\
\and
Institute of Natural Sciences, Shanghai Jiao Tong University, China\\
\and
School of Mathematical and Physical Sciences, Macquarie University, Australia
}
\maketitle              
\begin{abstract}
Large language models (LLMs) have garnered considerable attention for their proficiency in tackling intricate tasks, particularly leveraging their capacities for zero-shot and in-context learning. However, their utility has been predominantly restricted to general tasks due to an absence of domain-specific knowledge. This constraint becomes particularly pertinent in the realm of protein engineering, where specialized expertise is required for tasks such as protein function prediction, protein evolution analysis, and protein design, with a level of specialization that existing LLMs cannot furnish. In response to this challenge, we introduce \textsc{ProteinEngine}, a human-centered platform aimed at amplifying the capabilities of LLMs in protein engineering by seamlessly integrating a comprehensive range of relevant tools, packages, and software via API calls. Uniquely, \textsc{ProteinEngine} assigns three distinct roles to LLMs, facilitating efficient task delegation, specialized task resolution, and effective communication of results. This design fosters high extensibility and promotes the smooth incorporation of new algorithms, models, and features for future development. Extensive user studies, involving participants from both the AI and protein engineering communities across academia and industry, consistently validate the superiority of \textsc{ProteinEngine} in augmenting the reliability and precision of deep learning in protein engineering tasks. Consequently, our findings highlight the potential of \textsc{ProteinEngine} to bride the disconnected tools for future research in the protein engineering domain.

\keywords{Deep Learning \and Large Language Model \and Protein Design \and AI for Protein Design.}
\end{abstract}

\section{Introduction}
Large language models (LLMs) have achieved remarkable successes in solving complex tasks, showcasing their zero-shot learning capabilities~\cite{zhao2023survey}.
However, the effectiveness of these models tends to plateau when faced with more specialized tasks due to their inability to access domain-specific knowledge or utilize specialized tools tailored for certain applications. 
This limitation becomes glaringly apparent in the context of protein engineering tasks.
Although LLMs have been explored for specific tasks within the protein engineering domain such as protein structure prediction~\cite{lin2023evolutionary}, protein evolution analysis, or de novo protein design~\cite{ferruz2022protgpt2,madani2023progen}, their application often demands significant alterations to the model architecture.
These modifications, coupled with the need for learning from scratch using domain-specific datasets, present two-fold challenges. 
First, this approach underutilizes the capabilities of well-trained foundation models, given their robust pre-existing knowledge base. 
Second, the process of specialization often leads to the loss of the model's conversational abilities, a key feature that makes LLMs versatile and user-friendly.

The introduction of in-context learning capabilities in LLMs~\cite{incontext} has ushered potential solutions to the enduring challenge of domain knowledge scarcity~\cite{incontext}. 
In this new paradigm, the LLM acts as a centralized, cognitive-like system, which is capable of addressing domain-specific tasks by invoking relevant Application Programming Interfaces (APIs) or systems to bride the knowledge gap within the specific domain.
However, despite these advancements, most existing solutions tend to restrict the tools they incorporate to commonly used APIs such as calculators, calendars, and web searches, or relatively simple AI  models like text-to-image generation models.
In the context of protein engineering, both the task formulation and the APIs involved, as well as the AI models, manifest greater complexity. 
They are characterized by a diverse modality of input and a larger set of arguments, accentuating the need for more sophisticated and flexible systems. 
Consequently, further research and development are required to fully leverage the potential of LLMs in complex domains such as protein engineering.
To narrow the gap, we present \textsc{ProteinEngine}, a novel LLM system for protein engineering.
The major contributions are three-fold:
\begin{enumerate}
    \item[(1)] We introduce \textsc{ProteinEngine}, a human-centered platform to augment the capabilities of LLMs in tackling protein engineering tasks. 
    This is achieved by seamlessly integrating a comprehensive array of tools, packages, and software relevant to protein engineering, all accessible through APIs. 
    \item[(2)] We propose a role-playing framework, comprising AI Project Manager (AI-PM), AI Domain Expert (AI-DE), and AI Presenter (AI-Pr) modules, which facilitates efficient task delegation, promotes interdisciplinary integration, ensures dynamic adaptability, and enables effective communication of results respectively. 
    This design principle provides substantial flexibility, allowing for easy integration and extensibility with emerging AI protein-design models.
    \item[(3)] Through comprehensive user studies, we demonstrate the superior performance of \textsc{ProteinEngine} in not only enhancing the usability of currently disconnected protein engineering tools but also reducing the workload and learning difficulty across users with different backgrounds.
\end{enumerate}

\begin{figure}[t!]
    \centering
   \includegraphics[width=\textwidth]{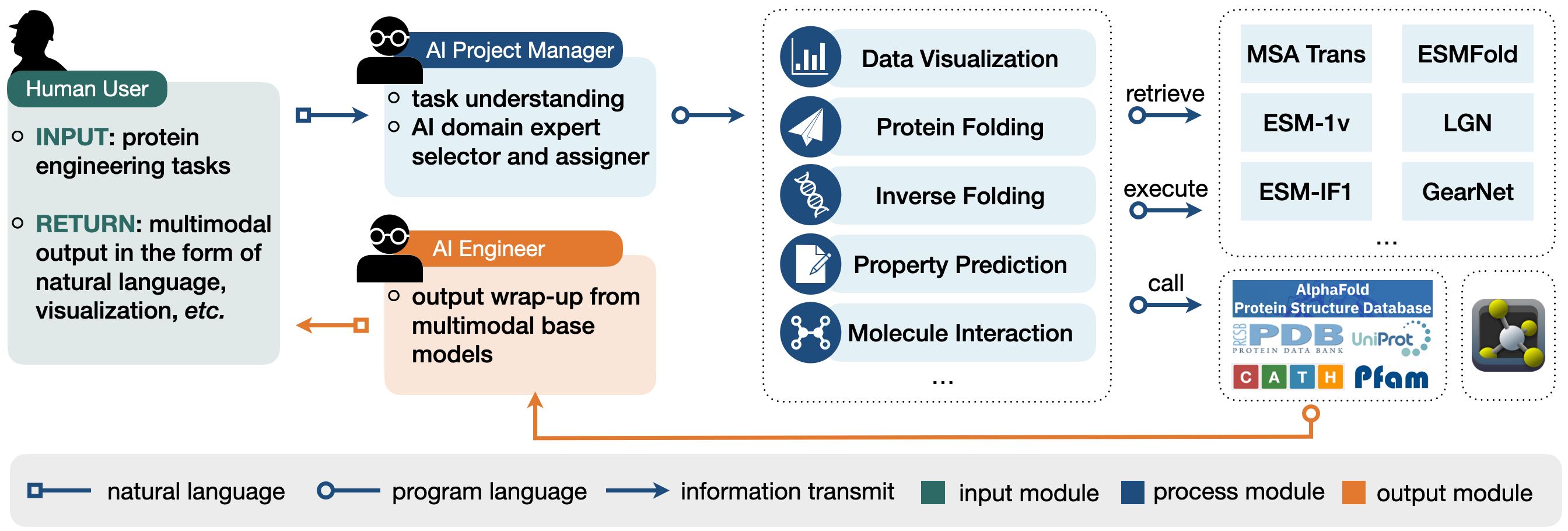}
    \caption{
    The overall framework of the proposed \textsc{ProteinEngine}, which incorporates three distinct roles, each assigned to a separate LLM.}
    \label{fig:platform}
\end{figure}

\section{\textsc{ProteinEngine}: LLM for Protein Engineering}
\paragraph{Method Overview}
Our proposed \textsc{ProteinEngine} is a human-centered system designed to augment the capabilities of existing LLMs to address a broader spectrum of intricate protein engineering tasks. 
Specifically, we assign three critical roles to the LLMs: the AI Project Manager (AI-PM), the AI Domain Expert (AI-DE), and the AI Presenter (AI-Pr). 
The overall pipeline illustrating these roles and their interactions is depicted in Figure~\ref{fig:platform}.
The AI-PM functions as the primary coordinator, interpreting user input expressed in natural language and ensuring all necessary inputs, configurations, arguments, and conditions are correctly provided. 
Subsequently, it breaks down a given complex task into smaller, more manageable sub-tasks, delegating them to the appropriate AI-DEs.
Then, we employ multiple AI-DEs within the platform to address the wide variety of challenges inherent in protein engineering and to facilitate future expansion. Each AI-DE specializes in a particular domain or category of tasks, ensuring a comprehensive coverage of the diverse aspects of protein engineering. 
During the inference stage, the AI-PM selects and assigns a subset of all AI-DEs to execute relevant APIs based on the nature of the task at hand.
Lastly, the AI-Pr is tasked with presenting the results, either unimodal or multimodal, generated by the AI-DEs to the user. 

\paragraph{AI Project Manager}
The LLM performing as the AI-PM acts as the primary interface, bridging the gap between the user and the underlying protein engineering tools within the \textsc{ProteinEngine} platform.
In its core role, the AI-PM is tasked with interpreting user input presented in natural language, discerning the context, and identifying the necessary tasks to be performed. 
Beyond this, the AI-PM ensures that all required inputs, arguments, configurations, and conditions are correctly provided. 
To accurately parse and deconstruct the user's query, the AI-PM uses in-context learning, a more efficient alternative to the computationally demanding process of LLM fine-tuning. 
By systematically decomposing complex tasks into smaller, manageable sub-tasks, the AI-PM ensures a thorough understanding of the user's requirements. 
Once the sub-tasks are defined, the AI-PM delegates them to the appropriate AI-DEs, taking into account their respective areas of specialization.

\paragraph{AI Domain Expert}
The AI-DE in the \textsc{ProteinEngine} is specifically designed to manage a distinct category of tasks pertaining to protein engineering. 
A team of multiple AI-DEs is assembled, with each expert equipped with the necessary domain-specific knowledge and tools to execute its designated tasks. 
To ensure the AI-DEs perform efficiently and adaptively, we have implemented a novel self-feedback communication loop mechanism between the AI-DE and the AI-PM. 
This autonomous mechanism, which operates without the need for human intervention, enables AI-DEs to progressively refine their understanding of new challenges that may arise during the execution process, and to seek assistance from their fellow AI-DEs, if required.
As a result, AI-DEs can dynamically adjust and respond to the evolving demands of the tasks, thereby maintaining a high degree of accuracy and effectiveness.

\paragraph{AI Presenter}
AI-Pr aggregates and presents the results generated by the AI-DEs in a clear, concise, and user-friendly manner, ensuring the user can easily interpret and utilize the generated insights, fostering a deeper understanding.
To effectively communicate the results to the user, the AI-Pr is capable of visualizing multimodal data, which includes, but is not limited to, images and textual data. 
This presentation is tailored to cater to different user preferences, and it enhances the comprehensibility of the data, enabling users to quickly grasp the key insights and outcomes delivered by the AI-DEs.

\begin{table}[t!]
\caption{The involved AI models and APIs for protein engineering in the proposed \textsc{ProteinEngine}.}
\label{tab:api}
\begin{center}
\resizebox{\textwidth}{!}{
    \begin{tabular}{lllll}
    \toprule
    \textbf{API} & \textbf{Functionality} & \textbf{Description} & \textbf{Input} & \textbf{Output} \\
    \midrule
    \textsc{AlphaFold 2} \cite{jumper2021highly} & protein folding & \makecell[l]{single-chain structure\\prediction with MSA} & protein sequence & \makecell[l]{atom-level 3D coordinates;\\residue-level pLDDT} \\
    \textsc{AlphaFold-Multimer} \cite{evans2022multimer}  & protein folding & \makecell[l]{multi-chain structure\\prediction with MSA} & protein sequence & \makecell[l]{atom-level 3D coordinates;\\residue-level pLDDT} \\
    \textsc{ESMFold} \cite{lin2023evolutionary} & protein folding & \makecell[l]{MLM-based structure\\prediction without MSA} & protein sequence & \makecell[l]{atom-level 3D coordinates;\\residue-level pLDDT} \\
    \textsc{MSA Transformer} \cite{rao2021msatranformer} & protein folding & \makecell[l]{single-chain structure\\prediction with MSA} & multiple sequence alignment & \makecell[l]{atom-level 3D coordinates;\\in \texttt{.pdb} format}
    \\ 
    \midrule
    \textsc{ESM-IF1} \cite{lin2023evolutionary} & inverse folding & \makecell[l]{single-site mutation \\ Transformer-based} & protein sequence & de novo protein sequence\\ 
    \textsc{LGN} \cite{zhou2023accurate} & variant effect prediction & \makecell[l]{deep mutation\\ GNN based denoising} & protein graph & de novo protein sequence\\
    \midrule
    \textsc{Equidock} \cite{ganea2021equidock} & protein-target interaction & rigid-body docking & \makecell[l]{two protein structures \\ in \texttt{.pdb} format} & binding affinity score\\
    \textsc{EquiBind} \cite{stark2022equibind} & protein-target interaction & rigid-body docking & \makecell[l]{protein-ligand structure \\ in \texttt{.pdb} format} & \makecell[l]{protein-ligand interaction sites\\ binding affinity score}\\
    \textsc{DiffDock} \cite{corso2023diffdock} & protein-target interaction & rigid-body docking & \makecell[l]{antibody-antigen structures\\ in \texttt{.pdb} format} & bound structure of complex\\
    \textsc{Diffab} \cite{luo2022antigenspecific}  & protein target interaction & antibody-antigen interaction & \makecell[l]{antibody-antibody structures\\ in \texttt{.pdb} format} & \makecell[l]{bound structure of complex\\ binding affinity, epitope mapping} \\ 
    \midrule
    \textsc{ProtENN} \cite{bileschi2022using} & sequence generation & \makecell[l]{language-based model} & protein sequence and structure & de novo protein sequence with function\\ 
    \textsc{Progen} \cite{madani2023progen} & sequence generation & \makecell[l]{language-based model} & protein sequence & de novo protein sequence with function\\ 
    \textsc{Grade-IF} \cite{yi2024graph} & sequence generation & \makecell[l]{graph-based model} & protein sequence & de novo protein sequence\\ \midrule
    \textsc{GearNet} \cite{zhang2023protein} & property prediction & \makecell[l]{latent representation\\ of protein structure} & protein graph & function or structure label\\ 
    \textsc{DeepSol} \cite{khurana2018deepsol} & property prediction & \makecell[l]{solubility prediction} & protein sequence & solubility\\
    \midrule  
    \textsc{PyMOL} & protein visualization & \makecell[l]{visualize 3D conformation \\for a given protein molecule} & \texttt{.pdb} document & 3D visualization of the protein\\ 
    \textsc{VMD} & protein visualization & \makecell[l]{visualize 3D conformation \\for a given protein molecule} & \texttt{.pdb} document & 3D visualization of the protein \\ 
\midrule%
    \textsc{BioMedLM} & biomedical domain Q\&A & \makecell[l]{trained on biomedical literature\\ and clinical notes} & \makecell[l]{natural language on\\ biomedicine or healthcare} & \makecell[l]{answer questions as a\\ specialist in the field}\\ 
    \bottomrule
    \end{tabular}
}
\end{center}
\end{table}

\begin{figure}[t!]
    \centering
   \includegraphics[width=0.9\linewidth]{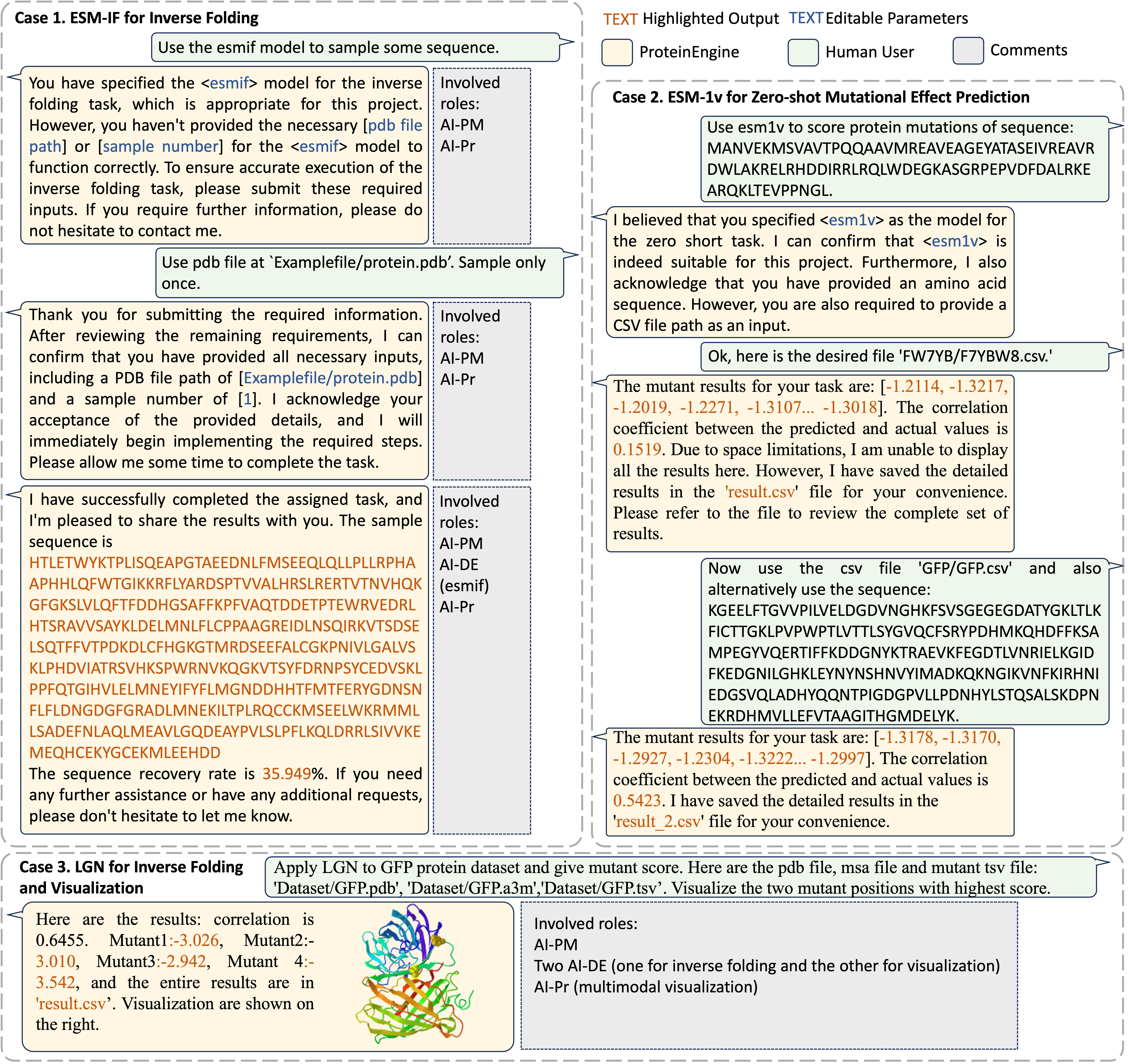}
    \caption{
    Three representative use case examples of the \textsc{ProteinEngine} in \texttt{user mode}, where only the absence of mandatory parameters will be requested to the user. 
    }
    \label{fig:chat}
\end{figure}

\paragraph{APIs for Protein Engineering}
The APIs integrated within our system, along with their corresponding task category taxonomy, are delineated in Table~\ref{tab:api}, which covers most of the task scenarios in protein design.
Each category is mapped to a specific AI-DE.
To provide a tangible understanding of the \textsc{ProteinEngine} in action, we illustrate its use through distinct case examples in Figure~\ref{fig:chat}, where different AI-DEs are involved in each case.
We illustrate typical examples of SOTA AI models in protein design which have been used in our \textsc{ProteinEngine} platform.

\section{User Study}

\begin{figure}[t!]
    \centering
   \includegraphics[width=0.75\linewidth]{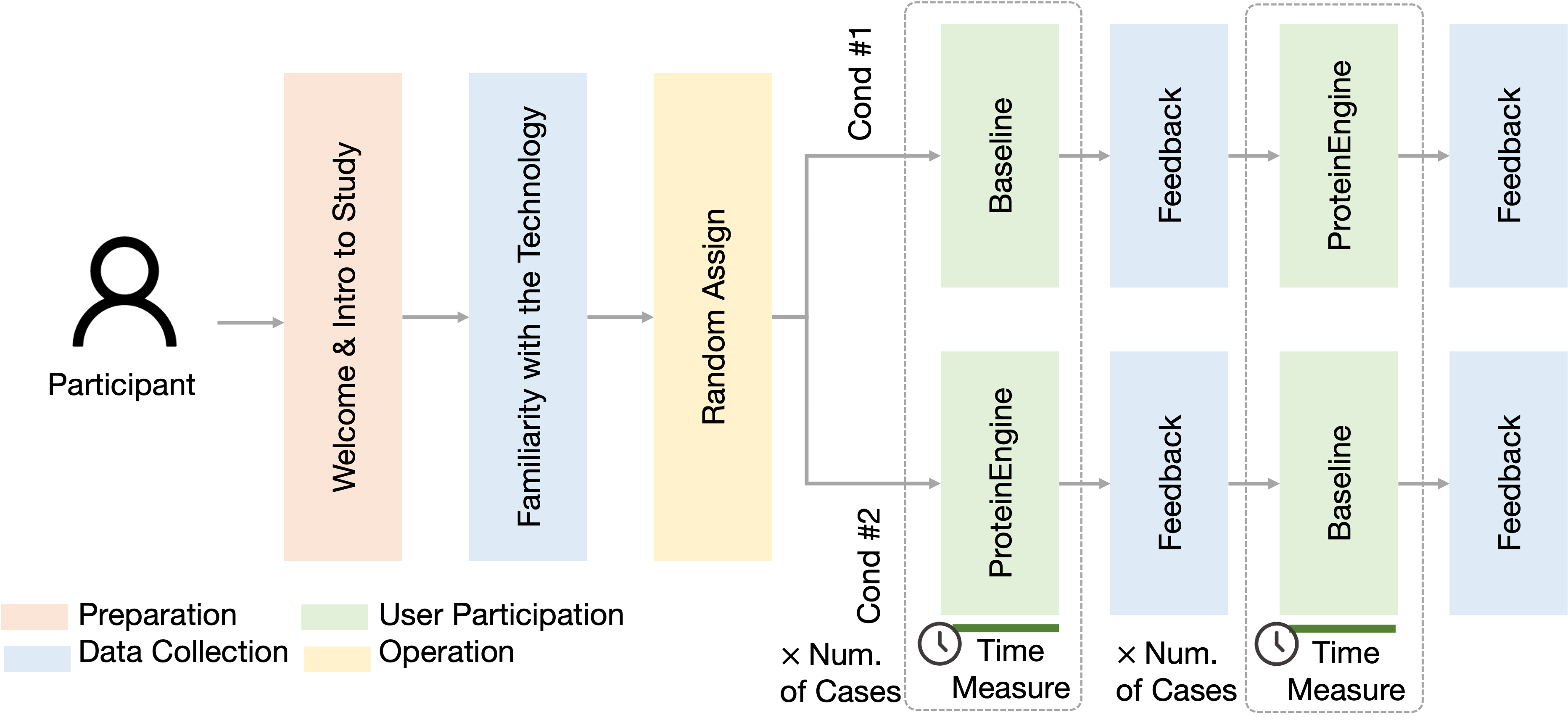}
    \caption{
    The overall flowchart of the user study. This includes preparation (participant recruitment and briefing), user participation operation (familiarization with the technology and random assignment to conditions), data collection (sequential tasks under different conditions with intermittent feedback). Each stage of the process is color-coded for ease of understanding.
    }
    \label{fig:userstudy}
\end{figure}

\paragraph{Hypothesis Formulation and Testing}
To evaluate the effectiveness of \textsc{ProteinEngine}, we conducted a user study focused on gauging its proficiency as an intuitive, human-centered system for protein engineering tasks. 
We employed hypothesis testing to quantitatively compare the performance of \textsc{ProteinEngine} against a baseline condition, focusing on task completion time, number of attempts, system usability, and the perceived workload.
In the baseline condition, participants employed traditional tools and methods, independent of \textsc{ProteinEngine}, such as executing Python scripts directly.
Therefore, our null hypotheses were formulated as follows:
\begin{enumerate}
    \item[(H1)] \textsc{ProteinEngine} does \textbf{not} reduce the time required for successful identification and execution of protein engineering models
    against the baseline.  
    \item[(H2)] \textsc{ProteinEngine} does \textbf{not} improve the accuracy in identifying and executing models for protein engineering tasks against \textsc{ProteinEngine} does improve the accuracy against the baseline. 
    \item[(H3)] \textsc{ProteinEngine} does \textbf{not} enhance the overall system usability for model identification and execution within protein engineering tasks against the baseline. 
    \item[(H4)] \textsc{ProteinEngine} does \textbf{not} decrease the workload required for the completion of protein engineering model identification and execution against the baseline.  
\end{enumerate}
Based on the hypotheses, we used a single-sided t-test to assess statistical significance.

\paragraph{Dependent Variables}
Statistical tests on these hypotheses involve collecting data on the dependent variables from the user study, specifically the task completion time, number of attempts, usability score, and  workload index, which are defined as follows.
\begin{itemize}[leftmargin=*]
    \item \textbf{Task Completion Time}:
    This objective, continuous variable measures the total time each participant takes to successfully complete a task under each condition. 
    \item \textbf{Number of Attempts}: Another objective, continuous variable records the total attempts a participant takes to successfully complete each task under each condition. 
    This variable is indicative of the accuracy of user actions. 
    \item \textbf{Usability Score}: This subjective, continuous variable is derived from the System Usability Scale (SUS) questionnaire~\cite{brooke1996sus}.
    The score reflects the perceived usability of the system. 
    \item \textbf{Workload Index}: This subjective, continuous variable, sourced from the NASA Task Load Index (NASA TLX) questionnaire~\cite{hart1988development}, assesses the perceived mental workload across six dimensions: mental demand, physical demand, temporal demand, effort, performance, and frustration level. 
\end{itemize}

\paragraph{Independent Variables}
The primary independent variable is the \textbf{Condition} under which participants perform the protein engineering tasks, either the baseline or the \textsc{ProteinEngine}.
Additionally, we consider potential confounding independent variables that could influence our study outcomes, including:
\begin{itemize}[leftmargin=*]
    \item \textbf{Participant Background}: These categorical variables encapsulate information about each participant's professional role and affiliations (academia or industry). 
    This information could offer insights into a user's likely background knowledge and potential biases or preferences when using the interface.
    \item \textbf{Familiarity with Technology}: 
    These numerical variables represent the degree of each participant's familiarity with protein engineering tasks, Python programming language, AI models, and the intersection of these areas.
    The level of familiarity could potentially influence the ease with which participants adapt to the \textsc{ProteinEngine}, and thus might impact the measurements of variables like task completion time, number of attempts, and perceived usability and workload.
\end{itemize}

\paragraph{User Study Design}
The user study workflow, shown in Figure~\ref{fig:userstudy}, consisted of several steps. 
First, participants received an introductory tutorial that provided information about the study's background, motivation, and procedures. Next, participants completed a preliminary questionnaire that assessed their background knowledge and familiarity with AI, protein engineering, and their interdisciplinary overlap.
The study followed a between-subjects design, comparing participants exposed to two conditions: a baseline condition and the \textsc{ProteinEngine} condition. To mitigate learning effects, the order in which participants encountered these conditions was randomized.
To control for potential effects stemming from participants' background and familiarity with technology, all participants completed the same set of activities under both conditions (baseline and \textsc{ProteinEngine}). This approach provided paired data for analysis.
Participants were assigned a series of six distinct protein engineering tasks under each experimental condition. These tasks included protein folding, inverse protein folding, and protein mutation prediction.
During the task completion process, we carefully recorded the total time taken for each task and the number of attempts required for successful execution.
After completing the tasks in either the baseline or \textsc{ProteinEngine} condition, participants were asked to fill out a questionnaire. This questionnaire aimed to assess their subjective impressions of the system's usability and their perceived workload during task completion.
To ensure a valid comparison of user experiences, identical questionnaires were administered. 
Data was collected using Google Sheets, with all questions being mandatory to prevent missing data. Incomplete data from participants who withdrew or failed to complete tasks were excluded from the analysis.

\paragraph{Participants Recruitment}
We strategically planned the recruitment of volunteer participants to encompass a wide range of potential users.
This design aimed to test the versatility and broad applicability of our proposed method across various user groups. 
Our participant cohort consisted of volunteers from both the AI and biological communities, spanning both academic and industrial fields. 
This diverse group, including students, AI researchers, lab technicians, and biologists, allows for a comprehensive evaluation of \textsc{ProteinEngine}'s functionality across multiple user profiles.

\paragraph{Implementations}
For the LLM in \textsc{ProteinEngine}, we chose the \texttt{gpt-3.5-turbo}.
As the most advanced model in the \textsc{GPT-3.5} series, this version provides robust capabilities and superior performance suitable for our application.

\paragraph{Collected Data}
\begin{figure}[b!]
    \centering
\includegraphics[width=0.7\linewidth]{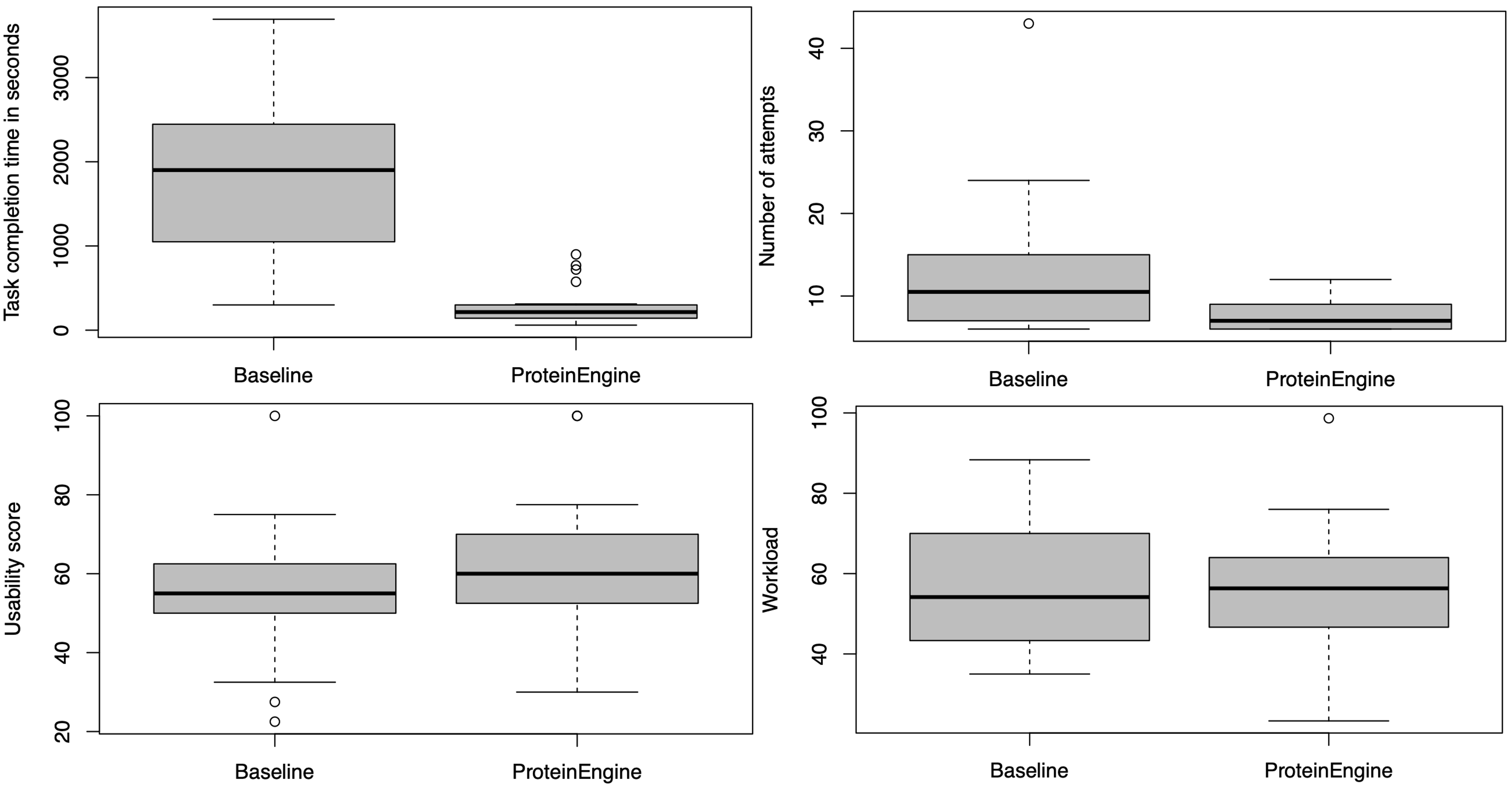}
    \caption{
    Boxplots for the four variables.
    }
    \label{fig:boxplot}
\end{figure}

The boxplots illustrating the distribution of our four key variables, namely Task Completion Time, Number of Attempts, Usability Score, and Workload Index, can be found in Figure~\ref{fig:boxplot}. Each box plot provides a visual summary of the minimum, first quartile (Q1), median (Q2), third quartile (Q3), and maximum values for these variables. The box represents the interquartile range (IQR) from Q1 to Q3, the line inside the box denotes the median, and the whiskers extend to show the range of the data within 1.5 times the IQR. Observations beyond this range are considered outliers and are represented as individual points.
%
%
On average it take 36.40 minutes to complete the user study for each participant.

\paragraph{Hypothesis Testing}
We applied the hypothesis testing to ascertain whether the performance differences observed between the two conditions i.e., baseline and \textsc{ProteinEngine} were statistically significant. 
The differences under consideration, represented as $d_i$, were computed by subtracting the \textsc{ProteinEngine} measurements from the baseline measurements. Under the null hypothesis, where both platforms have an equivalent effect, these differences should follow a distribution centered around zero, i.e., $\mu_d = 0$.
Our final dataset for hypothesis testing comprised $n = 26$ samples. 
We formulated our null and alternative hypotheses as follows:
$H_0: \mu_d = 0 \quad \texttt{against} \quad H_1: \mu_d > 0$.
This holds for testing hypotheses H1, H2, and H4. For testing H3, the alternative hypothesis is $\mu_d<0$.
Let $\overline{d}$ and $s_d$ represent the sample mean and sample standard deviation of the observed differences, respectively. Given these parameters, the sampling distribution for the test statistic follows a $t$ distribution with degrees of freedom $n-1$. This means that, under null hypothesis $H_0$, $\tau = \frac{\overline{d}}{s_d/\sqrt{n}} \sim t_{n-1}$.
%
Table \ref{tab:hypothesis} compiles our hypothesis testing results. Of note, no significant difference was found for the workload measure between the two conditions.

\begin{table}[t!]
\centering
\caption{Null hypothesis testing results.}
\label{tab:hypothesis}
\begin{tabular}{ c | c c c}
\hline
Hypothesis  &  Observed test statistics  & P-value & Reject null hypothesis \\ 
 \hline 
 H1 &  $7.7012$ & $2.335\times10^{-8}$ & Yes\\
 H2 & $3.1944$ & $1.884\times10^{-3}$ & Yes \\  
 H3 & $-2.4162$ & $1.166\times10^{-2}$ & Yes \\
 H4 & $0.74029$ & $2.330\times10^{-1}$ & No \\
 \hline 
\end{tabular}
\end{table}

\paragraph{Results}
Our study encompassed a total of 45 participants.
For hypothesis testing, we employed the paired two-sample t-test (effectively a one-sample, one-sided t-test on the difference) at a 5\% significance level for the four variables under two conditions, baseline and \textsc{ProteinEngine}.
The results allowed us to reject three of the null hypotheses, thereby highlighting the superior performance of \textsc{ProteinEngine} in facilitating protein engineering tasks.

\section{Conclusion}
In this work, we presented \textsc{ProteinEngine}, a groundbreaking platform that amplifies the capabilities of LLMs in the realm of protein engineering.
This platform's human-centered design greatly eases the learning curve traditionally associated with specialized tools, thereby making protein engineering tasks more accessible. 
By integrating advanced LLMs with domain-specific expertise, \textsc{ProteinEngine} marks a significant leap forward in the application of AI to protein engineering, showing great potential to accelerate scientific discoveries and spur innovation.
As we continue to develop and refine \textsc{ProteinEngine}, it is critical to emphasize the importance of responsible use and rigorous validation. 
Therefore, the development of comprehensive ethical guidelines and robust validation protocols is a key direction for future work. 

\bibliographystyle{plain}
\bibliography{reference}

\end{document}